\documentclass[smallextended]{svjour3}       
\usepackage{graphicx}
\begin{document}

\title{CPT analysis with top physics
}


\author{ Jose A. R. Cembranos
}


\institute{
Departamento de F\'{\i}sica Te\'orica I, \\
 Universidad Complutense de Madrid, \\
 E-28040 Madrid, Spain \\
              Tel.: +34-91-3948501\\
              Fax: +34-91-3945197\\
              \email{cembra@fis.ucm.es}           
}

\date{}

\maketitle

\begin{abstract}
We discuss the possibility of observing CPT violation from top anti-top production in hadronic colliders. We study a general approach by analyzing constraints on the mass difference between the top and anti-top quarks. We present current bounds from Tevatron data, and comment on the prospects for improving these bounds at the LHC and the ILC.
\end{abstract}

\section{Introduction}
\label{intro}

One of the most fundamental questions of theoretical science is related to the symmetries that underlay the laws of physics. In this work, we will study discrete symmetries, such as the charge conjugation symmetry $C$,  parity  $P$,  and the time reversal symmetry $T$. Other discrete symmetries are defined as their products. For instance, $CP$ is the product of the charge conjugation and parity and $CPT$ is the product of $CP$ and $T$.

In the past, laws of physics were assumed to conserve $C$, $P$ and $T$ , but a large number of experiments have contradicted this hypothesis. In fact, $C$ and $P$ are maximally violated in weak interactions \cite{P} and the neutral kaon system has shown evidences for the non conservation of either $CP$ \cite{CP} and $T$ \cite{T}. However there is not evidence of $CPT$ violation in any experiment, and on the contrary, there are important tests that constrain the amount of non conservation of $CPT$ in various sectors of the standard model of particles and interactions.

\begin{figure}[ht]
 \centerline{
    \includegraphics[width=0.8\textwidth,clip]{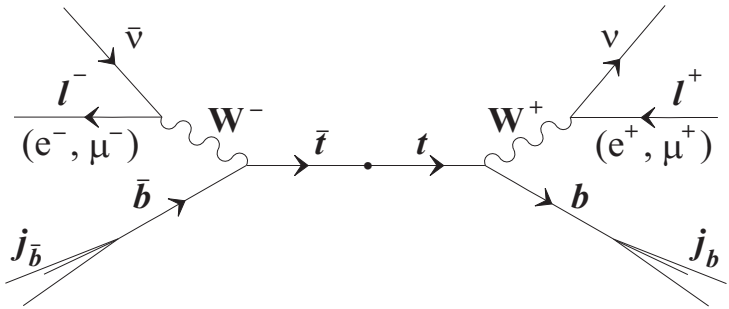}}
  \caption{Schematics of the top and anti-top decays in the dilepton channel.}
  \label{dileptonSch}
\end{figure}

A large number of models have been proposed in literature where $CPT$ violations can be accommodated, as the
standard--model extension associated with spontaneous breaking of the Lorentz symmetry in the string theory \cite{SME};
spacetime foam models motivated by quantum--gravity (QG) \cite{SF};
deformations of special relativity \cite{DSR} or modified gravity \cite{MG}; non-local models  originated from string theory \cite{NL};
or Lorentz symmetry breaking in extra dimensional models \cite{XD}.

In particular, extra dimensional holographic models are motivated as solution of the hierarchy problem.
Within these  models, the Higgs boson is a light composite pseudo Nambu-Goldstone boson.
The rest of the standard model particles are fundamentally elementary fields, decoupled to the new strong sector,
except the top quark, whose large mass is related to an important exposition to new physics.
In this framework, it is particularly interesting to search for $CPT$ violation in the top sector.
This analysis is independent and complementary to other studies in collider experiments \cite{Coll} or
astrophysical observations \cite{isearches} associated with the phenomenology of extra dimensional models.
The first bounds on $CPT$ violation within the top sector were published in 2008 \cite{CRT}, and they have experienced
an important development. In this work we will discuss the progress in this field and the best
direction for further improving.

\section{Hadronic Colliders}

$CPT$ conservation implies the same masses and lifetimes for particles and antiparticles. Any mass difference between a particle and its antiparticle is unambiguous evidence of $CPT$ violation. Here we  will focus on the measurement of the difference between the top and anti-top particles. The quantity $R_{CPT}(t)\equiv 2(m_t-m_{\bar t})/(m_t+m_{\bar t})$ is a useful dimensionless estimator of such a difference \cite{CRT}.

\begin{figure}[ht]
 \centerline{
    \includegraphics[width=0.8\textwidth,clip]{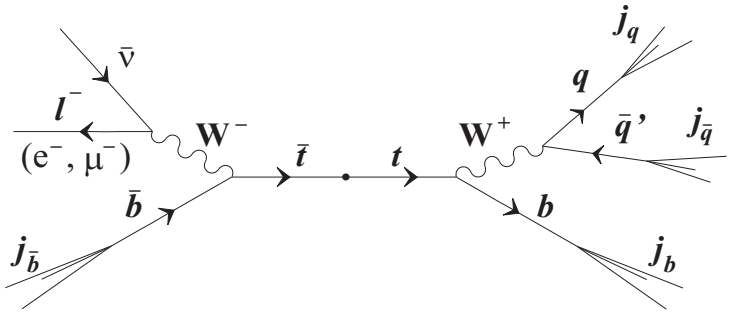}}
  \caption{Schematic example of the top and anti-top decays in the lepton plus jets channel.}
  \label{lplusjetsSch}
\end{figure}

\begin{figure}[ht]
    \begin{center}
        \includegraphics[width=0.8\textwidth,clip]{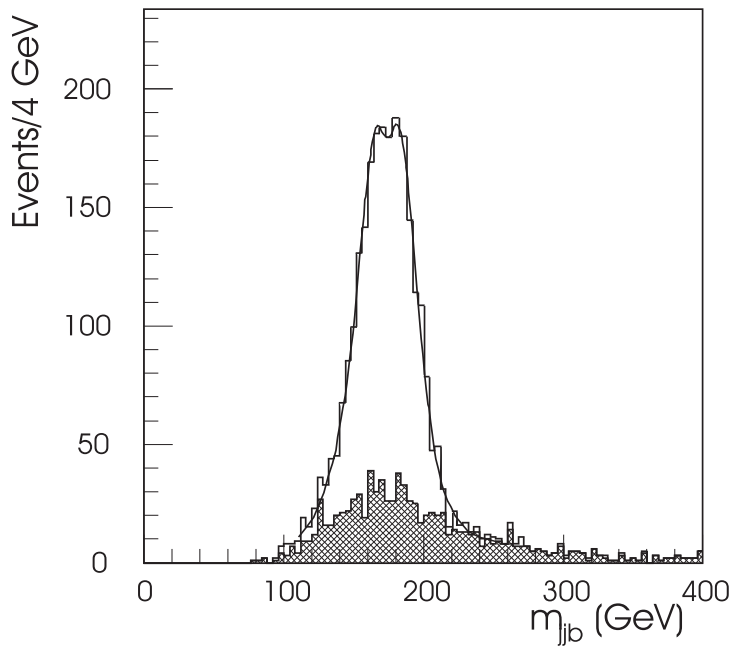}
    \end{center}
    \caption{Simulated invariant mass $m_{jjb}$ distribution for $R_{CPT}(t)=0.07$,
    i.e. $|m_t-m_{\bar t}|\simeq 12$ GeV. The number of events per bin is evaluated
    as the sum of the events produced either with top (with mass $m_t$) or anti-top
    (with a different mass $m_{\bar t}$). The shapes of these independent signals
    are assumed to have the standard invariant mass $m_{jjb}$ distribution from full simulation
    for the LHC for the lepton plus jets channel in top anti-top production \cite{CRT}.}
    \label{lhcmjjb}
\end{figure}

There are different analyses that can be performed at hadronic colliders. We will discuss top anti-top production identified in different channels (Figures \ref{dileptonSch} and \ref{lplusjetsSch}). We start with the study of the di-lepton channel, where the W bosons decay leptonically (see Figure \ref{dileptonSch}). We can reconstruct the top or anti-top mass by using the invariant mass associated to the lepton and b quark coming from the decay of the top or anti-top. The mass distribution from data coming from top and anti-top decays should have two different peaks if the $CPT$ violation is large enough, as it is shown by Figure \ref{lhcmjjb}. The first constraints on $R_{CPT}$ used the Tevatron data accumulated at Fermilab from 1992 through 1995 \cite{CDFdilept1}. The analysis performed by CDF by using this technique is consistent with only one peak, and the bound $|R_{CPT}(t)|<0.13$ was obtained at the 95\% c.l. \cite{CRT}. However, the same work found that a more constraining bound was provided by the lepton plus jet channel, in which one of the W bosons decays leptonically whereas the other one decays hadronically (Figure \ref{lplusjetsSch}). An analogous analysis to that of the di-lepton channel was done by reconstructing the masses with the invariant mass $m_{jjb}$ associated to the hadronic decay. Combining the CDF \cite{CDFl+jets} and DO data \cite{D0l+jets}, the constraint was $R_{CPT}(t) < 0.10$ \cite{CRT}.

Afterwards, the D0 Collaboration studied the top-antitop quark mass difference by using the matrix element technique \cite{ET}. Its first result,
with 1 fb$^{-1}$ of Run II integrated luminosity, implied an important improvement. Its analysis did not only constraint more efficiently the absolute value of the mass difference between top and anti-top, but it was sensitive to its sign:  $m_t-m_{\bar t}=3.8 \pm 3.4\,(\rm stat.) \pm 1.2\,(\rm syst.)$ GeV \cite{Abazov:2009xq}. A more updated work, with a total of 3.6 fb$^{-1}$ integrated luminosity, has obtained the present more constraining bound on $CPT$ violation in top physics: $m_t-m_{\bar t}=0.8\pm 1.8 (\rm{stat})\pm 0.5~(\rm{syst})$ GeV \cite{Abazov:2011ch}. Between both analysis, the CDF collaboration reported also a measurement of $m_t-m_{\bar t}=-3.3 \pm 1.4\,(\rm stat.) \pm 1.0\,(\rm syst.)$ GeV based on 5.6 fb$^{-1}$ of Run II data by using a template technique \cite{Aaltonen:2011wr}.

\section{Linear Colliders}

The same analyses can be performed with the International Linear Collider. There are fewer studies about the determination of the top mass through top anti-top quark production, but  the statistical uncertainties will increase while the  systematic errors can be reduced \cite{Biernacik:2003xv}.  The systematic ones dominate, at least in a conservative approach, and this fact leads to a small improvement of the sensitivity as compared to the LHC. In any case, the most promising study is associated with the threshold scan analysis for the production of top anti-top production in linear colliders since it is extremely sensitive to the top quark mass. The potential improvement could be of up to two orders of magnitude with respect to present measurements.

\begin{figure}[ht]
    \begin{center}
        \includegraphics[width=0.8\textwidth,clip]{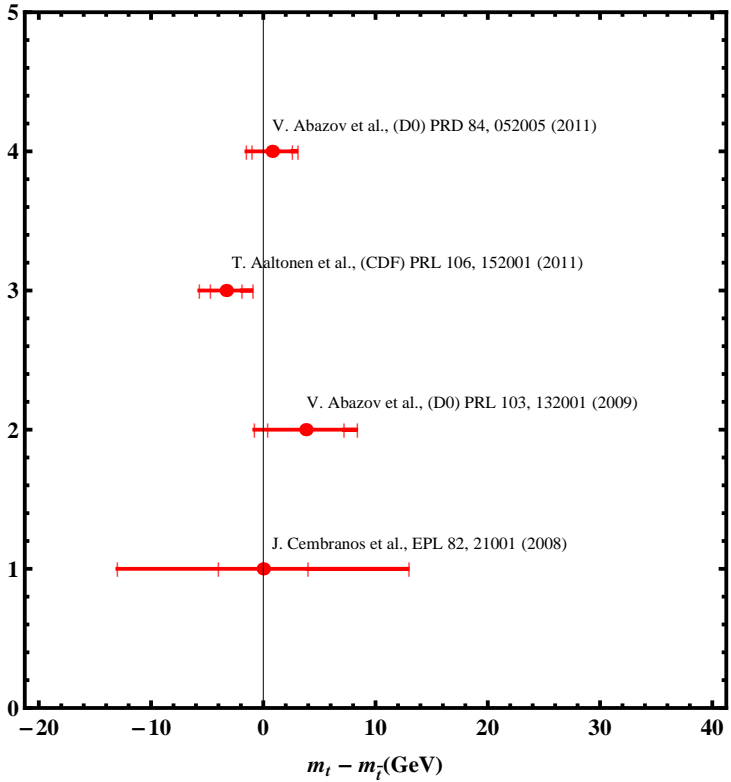}
    \end{center}
    \caption{Evolution of the constraints on the mass difference between top
    and anti-top quarks. The present improvement with respect to the first analysis in
    2008 \cite{CRT} is one order of magnitude, due fundamentally to a better control
    on systematic uncertainties.}
    \label{reswref}
\end{figure}

\section{CONCLUSIONS}

$CPT$ symmetry is guaranteed by the $CPT$ theorem based on three very fundamental assumptions: any local theory, which is invariant under Lorentz transformations and defined by a Hermitian Hamiltonian conserves $CPT$ \cite{local}. However, as we have commented, different models can produce $CPT$ violation. It is interesting to search for this violation in the frontiers of the standard model such as top physics \cite{toprev}. In the last years, an important improvement on the development of different techniques for measuring the mass difference between top and anti-top quarks has taken place. As Fig. \ref{reswref} summarizes, it has meant unprecedented progress in constraining $CPT$ violation within this sector.

\begin{acknowledgements}
This work has been supported by MICINN (Spain) projects numbers FIS2011-23000, FPA2011-27853-C02-01 and Consolider-Ingenio MULTIDARK CSD2009-00064.
JARC would like to thank the kind hospitality of ACGC/University of Cape Town while elaborating the manuscript.
\end{acknowledgements}

{\bf Note added:} CMS Collaboration published the first analysis with LHC data on the mass difference between top and anti-top as this contribution was being prepared \cite{Chatrchyan:2012ub}. It is based on almost 5 fb$^{-1}$ of integrated luminosity and it takes into account events with a lepton and at least four jets in the final state. The result: $m_t-m_{\bar t}=-0.44 \pm 0.46\,(\rm stat.) \pm 0.27\,(\rm syst.)$ GeV, shows another significant improvement with respect to the Tevatron analyses reported in this contribution.

\end{document}